\documentclass[12pt]{article}

\usepackage{amsmath,amsfonts,amssymb}

\def\he{\hat{e}}

\def\be{\begin{equation}}

\def\ee{\end{equation}}

\def\bea{\begin{eqnarray}}

\def\eea{\end{eqnarray}}

\def\bh{\bar{h}}

\def\mH{\mathcal{H}}

\def\hv{\hat{v}}

\newcommand{\mL}{\mathcal{L}}

\def\pb #1{\left\{#1\right\}}

\begin{document}

\begin{titlepage}

\vskip 0.4 cm

\begin{center}
{\Large{ \bf Relativistic and Newton-Cartan Particle in de Broglie-Bohm Theory
}}

\vspace{1em}  Josef Kluso\v{n}$\,^1$
\footnote{Email address:
 klu@physics.muni.cz}\\
\vspace{1em} $^1$\textit{Department of Theoretical Physics and
Astrophysics, Faculty
of Science,\\
Masaryk University, Kotl\'a\v{r}sk\'a 2, 611 37, Brno, Czech Republic}\\

%
%

\vskip 0.8cm

\end{center}

\begin{abstract}
This paper is devoted to the analysis of massive particle in
general and Newton-Cartan Background in de Broglie-Bohm Theory. We find
classical and quantum version of Hamilton-Jacobi equations and
find their relations to wave equations. We also discuss fundamental
difference between classical and quantum description of these
two systems.

\end{abstract}

\bigskip

\end{titlepage}

\newpage

\section{Introduction }
De Broglie-Bohm Mechanics also known as pilot-wave formulation was formulated by D. Bohm in \cite{Bohm:1951xw,Bohm:1951xx} based on previous
work of L. de Broglie
\cite{Broglie1,Broglie2} who suggested that particle trajectories are guided by waves. In this interpretation, individual quantum system is formed by a point particle \emph{and} a guiding wave. This is in sharp contrast with orthodox or Copenhagen interpretation of quantum mechanics, where an individual quantum system is simultaneously both a wave and particle. This system exhibits its particle or wave origin in dependence on the experiments. Say differently, in orthodox formulation of quantum mechanics does not make sense to speak about  particle's position  until its position is measured while in de Broglie-Bohm mechanics particle still has its own trajectory even if its form can differ from the classical one. It turns out that de Broglie-Bohm mechanics is,
in words of John Bell \cite{Bell:1987hh} "Bohmian
mechanics is a correct interpretation of quantum phenomena that exactly reproduce the predictions of the orthodox interpretation! "
\footnote{For review and extensive list of references, see
\cite{Berndl:1995ze,Tumulka:2018vot,Holland,Bohm}.}.
This theory  claims, that no rule of the quantum theory is broken when we include a single trajectory to specify the state of a system more finely than does the wave function. There is however an important point that whenever we want to calculate the trajectory we \emph{firstly} have to know the wave function \emph{whose determination is independent} on the path. This is main
objection to the de Broglie-Bohm theory since the concept of trajectory of particle can be interpreted as an redundant addendum and as an interpretative element rather than an intrinsic ingredient of quantum theory.

However there could be deeper reason why we should consider trajectory of particle as well defined physical object. The clue to this problem is in the similarity between the Schr\r{o}dinger equation in the position representation suitable expressed as two real equations. In more details, we can always write
the wave function in the form $\psi=\sqrt{\rho}e^{iS}$, where $\rho$ and $S$ are two real functions. Then it can be shown that the equation of motion for $S$ has the form of the quantum version of Hamilton-Jacobi (HJ) equation while the equation of motion for $\rho$ corresponds to the equation of continuity.
This is very nice picture and one can ask the question whether similar analysis can be extended to another situations. In fact, the goal of this paper is to perform similar analysis in case of relativistic massive particle in general background metric and gauge field and also to the case of particle in Newton-Cartan background
\footnote{For related work, see \cite{Nikolic:2012wj}.}. We show, following very nice analysis reviewed in
\cite{Oriols}, that even the classical relativistic particle and particle in Newton-Cartan background can be described with the help of wave function. We also show that fundamental difference between classical and quantum description is in the
the fact that the equations of motion that follow from the equations of motion for complex field in general background allow superposition of two solutions while in case of the classical description this is not possible. The same situation occurs in case
of particle in Newton-Cartan background. More precisely, we firstly determine an action for particle in Newton-Cartan background implementing null dimensional reduction
\cite{Festuccia:2016caf,Julia:1994bs}. Then we use the relation between conjugate momenta and partial derivative of the action in order to find
HJ equation for particle in Newton-Cartan background that has not been determined before. As the next step we determine an action for Schr\r{o}dinger field in Newton-Cartan background again using null dimensional reduction. Then we write this field using polar form and derive equations of motion where the first one correspond to the quantum HJ equation for particle in Newton-Cartan background while the second one corresponds to the equation of continuity. We also find an action for the complex scalar field that describes classical Newton-Cartan particle and we again show that the crucial difference between quantum and classical description is in the fact that linear combinations of two solutions of the classical equations of motion is not their solutions.

This paper is organized as follows. In the next section (\ref{second}) we review basic facts
about Hamilton-Jacobi theory which is central for formulation de Broglie-Bohm theory and relation between classical and quantum physics. In section (\ref{third}) we analyze relativistic massive particle in general background and find classical and quantum form of HJ equation for it. We also find Lagrangian for quantum relativistic particle. In section (\ref{fourth}) we
extend this analysis to the case of particle in Newton-Cartan background. Finally in conclusion (\ref{fifth}) we outline our results and suggest possible extension of this work.
\section{ Review of Hamilton-Jacobi Theory}\label{second}
In this section we briefly review basic facts about Hamilton-Jacobi theory, following
\cite{Oriols}, for more detailed treatment, see for example \cite{Goldstein}.
Let us consider an action defined as $S=\int d\lambda L(X,\dot{X})$ where
$X$ label position of particle in target space time and
 where $\dot{X}=\frac{dX}{d\lambda}$. All $X's$ depend on $\lambda$ which parameterizes world-line of particle. Let $\lambda_0$ and $\lambda_f$ are initial and final values of world-line parameters and let $X_0=X[\lambda_0], X_f=X[\lambda_f]$ are initial and final
positions of the particle. Classical trajectory, which we denote as $X_p[\lambda]$, is a trajectory that solves the equations of motion corresponds to the stationary value of the action functional
\begin{equation}
S[X,X_0,\lambda_0,X_f,\lambda_f]=
\int_{\lambda_0}^{\lambda_f}d\lambda L(X,\dot{X}) \ .
\end{equation}
More precisely, let $X_{np}(\lambda)=X_p(\lambda)+\delta X(\lambda)$ is a trajectory that differs from the physical one by small displacement parameterized by $\delta X$. We only demand that initial and final positions of these two trajectories are the same so that $\delta X(\lambda_0)=\delta X(\lambda_f)=0$. Then the action evaluated on the second trajectory is equal to
\begin{eqnarray}
& &S[X_{np},X_0,X_f]=\int_{\lambda_0}^{\lambda_f}
d\lambda L(X_p+\delta X,\dot{X}_p+\delta \dot{X}_p)=
\int_{\lambda_0}^{\lambda_f}d\lambda L(X_p,\dot{X}_p)+
\nonumber \\
& &+\int_{\lambda_0}^{\lambda_f}
d\lambda \left[\frac{\partial L}{\partial X}-\frac{d}{d\lambda}
\left[\frac{\partial L}{\partial \dot{X}}\right]\right]_{X=X_p}\delta X(\lambda)+
\left.\frac{\partial L}{\partial \dot{X}}\delta X\right|_{\lambda_0}^{\lambda_f}=
\nonumber \\
& &+\int_{\lambda_0}^{\lambda_f}
d\lambda \left[\frac{\partial L}{\partial X}-\frac{d}{d\lambda}
\left[\frac{\partial L}{\partial \dot{X}}\right]\right]_{X=X_p}\delta X(\lambda) \ ,
\nonumber \\
\end{eqnarray}
where we used the fact that
 $\delta X(\lambda_0)=\delta X(\lambda_f)=0$. Now the fact that
physical trajectory corresponds to the stationary point of the action means that
$S[X_{np}]-S[X_p]=0$ for any small variation $\delta X$ which can be also written as $\delta S=0$. As a result we derive Lagrangian equations of motion
\begin{equation}
\left[\frac{\partial L}{\partial X}-\frac{d}{d\lambda}
\left[\frac{\partial L}{\partial \dot{X}}\right]\right]_{X=X_p}=0 \ .
\end{equation}
Of course, this result  is well known. On the other hand it is less known
that the action formalism is more  powerful and allows more general analysis. For example,
we would like to analyze an ensemble of physical trajectories that differ in initial or final conditions. For that reason we define
\begin{equation}
S_{p-\delta \lambda}=S(X_{p-\delta \lambda})=
\int_{\lambda_0}^{\lambda_f+\delta \lambda}d \lambda
L(X_{p-\delta \lambda}(\lambda),\dot{X}_{p-\delta \lambda}[\lambda]) \ ,
\end{equation}
where $X_{p-\delta \lambda}[
\lambda]$ is physical trajectory that has identical initial and final positions as $X_p[\lambda]$ but taking a longer time $\lambda_f+\delta \lambda$.
Let us write this new physical trajectory as
$X_{p-\delta \lambda}(\lambda)=
X_p(\lambda)+\delta X(\lambda)$ for $\lambda\in [\lambda_0,\lambda_f]$. For $\lambda$ that is outsides from this interval we have
\begin{equation}\label{Xpexpansion}
X_{p-\delta \lambda}[\lambda_f+\delta \lambda]=
X_p[\lambda_f]+\dot{X}_p[
\lambda_f]\delta \lambda+\delta X[\lambda_f+\delta \lambda] \ .
\end{equation}
Since by definition we have
\begin{equation}\label{X-p}
X_{p-\delta \lambda}[\lambda_f+\delta \lambda]=X_p[\lambda_f]
\end{equation}
the equation (\ref{Xpexpansion}) implies
\begin{equation}
\delta X[\lambda_f+\delta \lambda]=-\dot{X}_p[\lambda_f]\delta \lambda \ .
\end{equation}
Then we can again write
\begin{eqnarray}\label{Sdeltap}
& &S_{p-\delta \lambda}=\int_{\lambda_0}^{\lambda_f+\delta \lambda}
d\lambda \left[L(X,\dot{X})_{X=X_p}+\left(\frac{\partial L}{\partial X}\delta X-
\frac{d}{d\lambda}\left[
\frac{\partial L}{\partial \dot{X}}\right]\right)_{X=X_p}\delta X\right]\nonumber \\
& &+\left[\frac{\partial L }{\partial \dot{X}}\delta X\right]_{
\lambda_0}^{\lambda_f+\delta \lambda} \ .  \nonumber \\
\end{eqnarray}
To proceed further we note that we have
\begin{eqnarray}
\int_{\lambda_0}^{\lambda_f+\delta \lambda}L(X)_{X=X_p}=
\int_{\lambda_0}^{\lambda_f}L(X)_{X=X_p}+\int_{\lambda_f}^{\lambda_f+\delta \lambda}L(X)_{X=X_p}
=S_p+L[X_p]\delta t \ . \nonumber \\
\end{eqnarray}
Finally note that the last term in (\ref{Sdeltap}) can be written as
\begin{eqnarray}
\left[\frac{\partial L}{\partial\dot{X}}\delta X\right]_{\lambda_0}^{\lambda_f+\delta \lambda}=
\frac{\partial L}{\partial \dot{X}}(\lambda_f+\delta \lambda)
\delta X(\lambda+\delta \lambda)=
-p(\lambda_f+\delta \lambda)\dot{X}(\lambda_f)
\nonumber \\
\end{eqnarray}
using (\ref{X-p}).
Finally, since $X_p$ is physical trajectory we find that the second term on the first line in (\ref{Sdeltap}) is zero. Putting these terms together we find
\begin{equation}
S_{p-\delta \lambda}=S_\lambda+L\delta \lambda-p\dot{X}\delta \lambda=S_p-H\delta \lambda
\end{equation}
that can be written as
\begin{equation}\label{partialSlambdaf}
\frac{\partial S(X_p(\lambda_f))}{\partial \lambda_f}=
-H(X_p(\lambda_f)) \ .
\end{equation}
As the next step we analyse the value of the action $S_{p-\delta X}$ when
we modify the final position $X_f+\delta X$ without modification of the initial
and final times.  The final position on the new physical trajectory $X_{p-\delta X}(\lambda)=X[\lambda]+\delta X[\lambda]$ means
\begin{equation}
\delta X[\lambda_0]=0 \ , \quad \delta X[\lambda_f]=\delta X_f \ .
\end{equation}
Let us now evaluate the action on this new trajectory
\begin{eqnarray}\label{SdeptaX}
& &S_{p-\delta X}=\int_{\lambda_0}^{\lambda_f}d\lambda L(X)|_{X=X_p+\delta X}=\nonumber \\
& &=\int_{\lambda_0}^{\lambda_f}d\lambda L[X_p]+
\left[\frac{\partial L}{\partial \dot{X}}\delta X\right]_{\lambda_0}^{\lambda_f}
+\int_{\lambda_0}^{\lambda_f}
d\lambda \left[\frac{\partial L}{\partial X}-\frac{d}{d\lambda}
\left[\frac{\partial L}{\partial \dot{X}}\right]\right]_{X=X_p}\delta X \ .
\end{eqnarray}
The first term on the second line in (\ref{SdeptaX})
is $S_p$. Further, the last expression on the second line in (\ref{SdeptaX})
vanishes  due to the fact that the physical trajectory solves the
equation of motion.
Finally the second term is equal to
\begin{eqnarray}
\left[\frac{\partial L}{\partial \dot{X}}\delta X\right]_{\lambda_0}^{\lambda_f}=
\left. \frac{\partial L}{\partial \dot{X}}\right|_{X=X_p}\delta X_f=p[\lambda_f]\delta X_f \ .
\nonumber \\
\end{eqnarray}
Then (\ref{SdeptaX}) can be written as
\begin{equation}\label{partSx}
\frac{\partial S}{\partial X_f}=\lim \frac{S_{p-\delta X}-S_p}{\delta X_f}=
p[\lambda_f] \ .
\end{equation}
In summary, the variation of the action with respect to final point is equal to momentum
of the trajectory at the final time. Then collecting all terms together we obtain
Hamilton-Jacobi equation
\begin{equation}
\frac{\partial S}{\partial \lambda}+H\left(X,\frac{\partial S}{\partial X},\lambda\right)=0 \ .
\end{equation}
\section{Relativistic  Classical and Quantum Particle}\label{third}
\subsection{Derivation of Relativistic Hamilton-Jacobi equation}
Let us consider a relativistic particle with the action
\begin{equation}\label{relpartaction}
S=-m\int d\lambda \sqrt{-g_{MN}\dot{X}^M\dot{X}^N}+q\int d\lambda A_M \dot{X}^M \ ,
\end{equation}
where $X^M,M=0,\dots,D-1$ label position of particle in target space-time
with general metric $g_{MN}(X)$ and gauge field $A_M$. Further, $m$ is the mass of the particle while $q$ is its charge. Now from the action (\ref{relpartaction}) we derive following conjugate momenta
\begin{equation}\label{relp}
p_M=m
\frac{g_{MN}\dot{X}^N}{\sqrt{-g_{MN}\dot{X}^M\dot{X}^N}}+qA_M \ .
\end{equation}
It is easy to see that the
 Hamiltonian is zero
\begin{equation}
H=p_M\dot{X}^M-L=0
\end{equation}
and hence Hamilton Jacobi equation implies $\frac{\partial S}{\partial \lambda}=0$. In other words in case of the relativistic theory there is no
explicit dependence of the action on parameter $\lambda$. In order to find
relativistic HJ equation we firstly use the definition of $p_M$ given in
(\ref{relp}) and the relation between partial derivative of action and conjugate momenta as was given in (\ref{partSx}). Explicitly, using (\ref{relp}) we derive
%
classical HJ equation for relativistic particle
\begin{equation}\label{relpartHJ}
(\partial_M S-qA_M)g^{MN}(\partial_MS-qA_M)+m^2=0 \ .
\nonumber \\
\end{equation}
\subsection{Ensemble of Trajectories}
The solution of the HJ equation is function $S$ that depends on space-time coordinates $x$. Then it is more natural to use it for the description of ensemble of trajectories where each trajectory differs by initial position and velocity. In order to describe
this fact we introduce $\rho$ as a function of space-time points that describe
distribution of the initial positions at time $\lambda_0$. These trajectories then evolve according to HJ equations. Further, let us define following particle four-velocity
\begin{equation}
u^M(\lambda)=\frac{\dot{X}^M}{\sqrt{-g_{MN}\dot{X}^M\dot{X}^N}} \ .
\end{equation}
To see that this is a correct form of the particle four-velocity let us presume that there is well defined relation between arbitrary parameter $\lambda$ and proper-time $\tau$ in the form $\lambda=\lambda(\tau)$. Then we can rewrite $u^M$ as
\begin{equation}
u^M=\frac{\frac{dX^M}{d\tau}\frac{d\tau}{d\lambda}}{
	\sqrt{-g_{MN}\frac{dX^M}{d\tau}\frac{dX^N}{d\tau}}\frac{d\tau}{d\lambda}}=
\frac{dX^M}{d\tau}
\end{equation}
since $\sqrt{-g_{MN}\frac{dX^M}{d\tau}\frac{dX^N}{d\tau}}=1$.  On the other hand we know that $u^M$ can be expressed with the help of the partial derivative of $S$ and external gauge field $A_M$ as
\begin{equation}
u^M=g^{MN}(\partial_N S-qA_N) \ .
\end{equation}
Hence it is natural to define current density in the form
\begin{equation}
J^M=\rho \sqrt{-g}g^{MN}(\partial_NS-qA_N) \
\end{equation}
which is current defined in each point of  space-time. This current is
conserved
\begin{equation}
\partial_M J^M=\partial_M (\sqrt{-g}\rho g^{MN}(\partial_NS-qA_N))=0 \
\end{equation}
which expresses the fact that   trajectory cannot be created or destroyed.
In summary, in case of the ensemble of trajectories of  classical relativistic particle we have following two equations
\begin{eqnarray}\label{classsreleqm}
& &\partial_M[\rho \sqrt{-g}g^{MN}(\partial_NS-qA_N)]=0
\ , \nonumber \\
& &(\partial_M S-qA_M)g^{MN}(\partial_NS-qA_N)+m^2=0 \ .
\nonumber \\
\end{eqnarray}
We see that even classical particle has description with the help of two fields
$S(x)$ and $\rho(x)$, where $S(x)$ is action function that solves classical relativistic Hamilton Jacobi equation while $\rho(x)$ describes a distribution of
trajectories from ensemble over space-time. It is also nice that
 these equations of motion can be derived from the action
\footnote{We work in units where $\hbar=1$.}
\begin{equation}\label{Iclassrel}
I_{class}=-\int d^Dx \sqrt{-g}(\rho g^{MN}(\partial_MS-qA_N)(\partial_MS-qA_M)+m^2\rho)
 \ .
\end{equation}
In fact, performing variation of the action (\ref{Iclassrel}) with respect to $\rho$ we obtain the second equation in (\ref{classsreleqm}) while performing
variation with respect to $S$ we obtain  first equation in (\ref{classsreleqm}).
As the next step we introduce complex field $\psi_{clas}=\sqrt{\rho}e^{iS}$ so that we can write
\begin{equation}
\rho=\psi_{clas} \psi^*_{clas} \ , \quad  S=\frac{1}{2i}\ln \frac{\psi_{clas}}{\psi^*_{clas}} \ .
\end{equation}
Then it is easy to see that the action (\ref{Iclassrel}) can
be written in the form
\begin{eqnarray}
I_{class}
&=&\int d^Dx \sqrt{-g}\left(-\frac{1}{2}D_M\psi (D_N\psi)^*g^{MN}+\right.
\nonumber \\
& & \left.+
\frac{\psi^*_{clas}}{4\psi_{clas}}
D_M\psi_{clas} D_N\psi_{clas} g^{MN}+
\frac{\psi_{clas}}{4\psi_{clas}^*}(D_M\psi_{clas})^*(D_N\psi^{clas})^*g^{MN}+m^2
\psi_{clas} \psi^*_{clas}\right) \ ,
\nonumber \\
\end{eqnarray}
where
\begin{equation}
D_M\psi_{clas}=\partial_M\psi_{clas}-iqA_M\psi_{clas} \ .
\end{equation}
Note that the equation of motion that follow from the action above takes the form
\begin{eqnarray}\label{eqkkclas}
& &	\frac{1}{2}D_M[\sqrt{-g}g^{MN}
	D_N\psi_{clas}]
+\frac{1}{4\psi_{clas}}D_M\psi_{clas}D_N\psi_{clas}g^{MN}\sqrt{-g}+m^2\psi_{clas}\sqrt{-g}-
\nonumber \\
&-&\frac{\psi_{clas}}{4(\psi_{clas})^*}
(D_M\psi_{clas})^*(D_N\psi_{clas})^*g^{MN}\sqrt{-g}
-\frac{1}{2}D_M\left[\frac{\psi_{clas}}{\psi_{clas}^*}
\sqrt{-g}g^{MN}(D_N\psi_{clas})^*\right]=0 \ .
\nonumber \\
\end{eqnarray}	
The crucial point of this equation is the lack of linearity of solutions. In fact, if $\psi^1_{clas}$ and $\psi^2_{clas}$ are  two solutions of the equation (\ref{eqkkclas})
then their linear combinations $a\psi_{clas}^1+b\psi_{clas}^2$, where $a,b$ are two complex numbers cannot be solutions of (\ref{eqkkclas}). This is fundamental difference between classical and quantum relativistic particle.
\subsection{Quantum Relativistic Particle}
Now we consider complex field action
\begin{equation}
I=-\int d^Dx \sqrt{-g}[g^{MN}D_M \Psi (D_N \Psi)^*+m^2\Psi \Psi^*] \ ,
\end{equation}
where $D_M\Psi=\partial_M\Psi-iqA_M\Psi$. The equation of motion derived
from this action has the form
\begin{equation}
D_M[\sqrt{-g}g^{MN}D_N\Psi]-m^2\Psi=0
\end{equation}
that are clearly linear which is fundamental difference between classical
and quantum description.

As the next step we introduce polar decomposition of  $\Psi$ as
\begin{equation}
\Psi=\sqrt{\rho}e^{iS} \ ,
\end{equation}
so that we have
\begin{equation}
D_M\Psi=\frac{1}{2\sqrt{\rho}}\partial_M\rho e^{iS}+i\sqrt{\rho}(\partial_M S-qA_M)e^{iS} \ .
\end{equation}
Then it is easy to see that the action has the form
\begin{eqnarray}
I=-\int d^Dx \sqrt{-g}\left(g^{MN}\frac{1}{4\rho}\partial_M\rho
\partial_N\rho+\rho  g^{MN}(\partial_M S-qA_M)(\partial_NS-qA_N)+m^2\rho \right)
\nonumber \\
\end{eqnarray}
with following equations of motion
\begin{eqnarray}\label{KGeqm}
& &-\sqrt{-g}\frac{1}{4\rho^2}g^{MN}\partial_M\rho\partial_N\rho
-\partial_M[\sqrt{-g}g^{MN}\frac{1}{4\rho}\partial_N\rho]+
\nonumber \\
& &+\sqrt{-g}g^{MN}(\partial_MS-qA_M)(\partial_N S-qA_N)+m^2\sqrt{-g}=0 \ ,
\nonumber \\
& &\partial_M[\sqrt{-g}g^{MN}\rho(\partial_NS-qA_N)]=0 \ .
\end{eqnarray}
Note that the first equation can be interpreted as the quantum relativistic
Hamilton-Jacobi equation when we write it in the form
\begin{eqnarray}
& &g^{MN}(\partial_M S-qA_M)(\partial_NS-qA_N)+m^2+Q=0 \ , \nonumber \\
& & Q=-\frac{1}{4\rho^2}g^{MN}\partial_M \rho \partial_N\rho-\frac{1}{\sqrt{-g}}
\partial_M[\sqrt{-g}g^{MN}\frac{1}{4\rho}\partial_N\rho]  \ .
\nonumber \\
\end{eqnarray}
Now we would like to ask a question what is a consequence of the quantum
potential $Q$ on the dynamics of quantum particle. Since $p_M=\partial_MS$ we can
rewrite the HJ equation into the form
\begin{equation}
(p_M-qA_M)g^{MN}(p_N-qA_N)+m^2+Q=0 \
\end{equation}
which is nothing else than the quantum version of Hamiltonian constraint. As
a consequence the Hamiltonian has the form
\begin{equation}
H_Q=\Gamma\left( (p_M-qA_M)g^{MN}(p_N-qA_N)+m^2+Q\right)
\ ,
\end{equation}
where $\Gamma$ is Lagrange multiplier corresponding to the Hamiltonian constraint.
Let us now derive corresponding Lagrangian. Using the Hamiltonian given above we obtain
\begin{equation}
\dot{X}^M=\pb{X^M,H_Q}=2\Gamma g^{MN}(p_N-qA_N)
\end{equation}
and it is easy to find
\begin{equation}
L_Q=p_M\dot{X}^M-H_Q=
\frac{1}{4\Gamma}\dot{X}^Mg_{MN}\dot{X}^N-\Gamma (m^2+Q) \ .
\end{equation}
Finally we can eliminate $\Gamma$ using corresponding equations of motion
\begin{equation}
-\frac{1}{4\Gamma^2}
\dot{X}^Mg_{MN}\dot{X}^N-(m^2+Q)
=0
\end{equation}
that has solution
\begin{equation}
\Gamma=\frac{\sqrt{-g_{MN}\dot{X}^M\dot{X}^N}}{2\sqrt{m^2+Q}} \ .
\end{equation}
Inserting back to the Lagrangian we derive the final form of the
Lagrangian for quantum relativistic particle in Bohmian mechanics
\begin{equation}
L_Q=-\frac{1}{2}\sqrt{m^2+Q}\sqrt{-g_{MN}\dot{X}^M\dot{X}^N}+qA_N\dot{X}^N \ .
\end{equation}
We mean that this is very remarkable result. In fact, solving the equations of motion that follow from this Lagrangian we derive quantum trajectories for relativistic particle where these trajectories are strongly affected by quantum potential $Q$ which depends on function $\rho$. It is further very interesting that the quantum potential appears in the relativistic Lagrangian under square root together with the rest mass of the particle. Of course, in order to determine quantum trajectories we have to know the form of $\rho$ that is determined by second equation in (\ref{KGeqm}) that, since it contains $S$ has to be solved together with the first equation in (\ref{KGeqm}).
\section{Particle in Newton-Cartan Background and Its Hamilton-Jacobi Equation}
\label{fourth}
In this section we find Hamilton-Jacobi equation for particle in Newton-Cartan  background. To do this we find an action for massive particle in this background with the help of null dimensional reduction, following
\cite{Festuccia:2016caf,Julia:1994bs}.
Explicitly, let us consider an action for massive particle in $D+1$ dimensions with the background metric $g_{AB}$ and the gauge field $A_A$ where $A=0,\dots,D$. The particle action has standard form
\begin{equation}\label{massparticleaction}
S=-T\int d\lambda \sqrt{-\gamma_{AB}\dot{X}^A\dot{X}^B}+q\int d\lambda A_M
\dot{x}^M \ ,
 \end{equation}
 where $T$ is the mass of particle and where $q$ is its charge. As the next
 step we determine Hamiltonian for this particle. From (\ref{massparticleaction})
we find following conjugate momenta
\begin{equation}
p_A=T\frac{\gamma_{AB}\dot{X}^B}{\sqrt{-\gamma_{AB}\dot{X}^A\dot{X}^B}}+q A_A
\end{equation}
that implies following Hamiltonian constraint
\begin{equation}
\mH_\tau=(p_A-qA_A) \gamma^{AB}(p_B-qA_B)+T^2\approx 0 \ .
\end{equation}
Let us now consider background metric which possesses a null
isometry that is generated by coordinates $\partial_u$
\begin{equation}
ds^2=\gamma_{AB}dx^A dx^B=2\tau_\mu dx^\mu(du-M_\nu dx^\nu)+h_{\mu\nu} dx^\mu dx^\nu \ , \quad \mu,\nu=0,1,\dots,D-1
\end{equation}
so that
\begin{equation}\label{gammametric}
\gamma_{\mu u}=\gamma_{u\mu}=\tau_\mu \ , \quad \gamma_{\mu\nu}=
h_{\mu\nu}-\tau_\mu M_\nu -\tau_\nu M_\mu\equiv \bh_{\mu\nu}
\
\end{equation}
and also
\begin{equation}
\sqrt{-\gamma}=e \ , \quad
e=\det \left(\tau_\mu, e_\mu^{ \ a}\right) \ .
\end{equation}
Note that the metric inverse to (\ref{gammametric}) has the form
\begin{equation}\label{gammametricinv}
\gamma^{uu}=2\Phi \ , \quad  \gamma^{u\mu}=-\hv^\mu \ , \quad  \gamma^{\mu\nu}=h^{\mu\nu} \ ,
\end{equation}
where we defined spatial vierbein $e^\mu_{ \ a}$ and $e_\mu^{ \ a} \ , a=1,\dots,D-1$ as
\footnote{We follow convention used in \cite{Festuccia:2016caf}.}
\begin{equation}
h_{\mu\nu}=e_\mu^{ \ a}e_\nu^{ \ b}\delta_{ab} \ , \quad
h^{\mu\nu}=e^\mu_{ \ a}e^\nu_{ \ b}\delta^{ab} \  , \quad
e^\mu_{ \ a}e_\mu^{ \ b}=\delta_a^b \ .
\end{equation}
We also introduce temporal vector $v^\mu$ that obeys the relation
\begin{equation}
v^\mu\tau_\mu=-1 \ , \quad v^\mu e_\mu^{ \ a}=0 \ , \quad
\tau_\mu e^\mu_{ \ a}=0 \ .
\end{equation}
Finally we introduced Galilean invariant objects $\hv^\mu, \he_\mu^{ \ a}$ and
$\Phi$ defined as
\begin{eqnarray}
\hv^\mu=v^\mu-h^{\mu\nu}M_\nu \ , \quad
\he_\mu^{ \ a}=e_\mu^{ \ a}-M_\nu e^\nu_{ \ b}\delta^{ba}
\tau_\mu \ , \quad
\Phi=-v^\mu M_\mu+\frac{1}{2}h^{\mu\nu}M_\mu M_\nu \ .
\nonumber \\
\end{eqnarray}
%
We further presume that the gauge field has the form $A_A=(A_u,A_\mu)=
(\varphi,\bar{A}_\mu-\varphi M_\mu)$.
Now with the help of this metric we perform dimensional reduction.
To do this we have to presume that all fields do not depend on $u$.
Then using the previous form of the Hamiltonian constraint and the form
of the background field we find that the Hamiltonian constraint has the form
\begin{eqnarray}
& &\mH_\tau=2(p_u-q\varphi)\Phi (p_u-q\varphi)
-2(p_u-q\varphi) \hv^\mu (p_\mu-q(\bar{A}_\mu-\varphi M_\mu))+\nonumber \\
& &+h^{\mu\nu}(p_\mu-q(\bar{A}_\mu-\varphi M_\mu))( p_\nu-q(\bar{A}_\nu-\varphi M_\nu))+T^2\approx 0 \ .
\nonumber \\
\end{eqnarray}
Let us now assume that the particle has constant momentum along $u-$direction
that we denote as $m$. This is certainly reasonable assumption due to the fact that  background fields do not depend on $u-$coordinate.
 We also see that it is possible to take the limit
$T\rightarrow 0$ at the Hamiltonian constraint. In the end we obtain
Hamiltonian constraint in the form
\begin{eqnarray}
& &\mH_\tau=2(m-q\varphi)\Phi (m-q\varphi)
-2(m-q\varphi) \hv^\mu (p_\mu-q(\bar{A}_\mu-\varphi M_\mu))+\nonumber \\
& &+h^{\mu\nu}(p_\mu-q(\bar{A}_\mu-\varphi M_\mu))( p_\nu-q(\bar{A}_\nu-\varphi M_\nu))\approx 0 \ , \quad
H=\Gamma \mH_\tau  \ .
\nonumber \\
\end{eqnarray}
In order to find corresponding Lagrangian we determine
equations of motion for $X^\mu$
\begin{equation}
\dot{X}^\mu=\pb{X^u,H}=\Gamma 2h^{\mu\nu}(p_\nu-q(\bar{A}_\nu-\varphi M_\nu))-2(K-q\varphi)\Gamma\hv^\mu \ ,
\end{equation}
where $\Gamma$ is Lagrange multiplier corresponding to the constraint $\mH_\tau\approx 0$. Then we obtain Lagrangian for particle in Newton-Cartan background in the form
\begin{eqnarray}\label{LNCparticle}
& &\mL=p_\mu \dot{X}^\mu-H=\nonumber \\
&=&\Gamma (p_\mu-q(\bar{A}_\mu-\varphi M_\mu)) h^{\mu\nu}(p_\nu-q(\bar{A}_\nu-
\varphi M_\nu))-
2(m-q\varphi)^2\Phi\Gamma+qA_\mu\partial_\lambda x^\mu \ . \nonumber \\
\end{eqnarray}
To proceed further we use the fact that
\begin{eqnarray}
& &\he_\mu^{ \ a}\dot{X}^\mu=2\Gamma e^\nu_{ \ c}\delta^{ca}(p_\nu-q(\bar{A}_\nu-
\varphi M_\nu))
\ ,  \nonumber \\
& &\dot{X}^\nu \he_\nu^{ \ a}\delta_{ab}\he_\mu^{ \ b}\dot{X}^\mu=
4\Gamma^2 (p_\mu-q(\bar{A}_\mu-\varphi M_\mu)) h^{\mu\nu}(p_\nu-q(\bar{A}_\nu-\varphi M_\nu))
\nonumber \\
\end{eqnarray}
so that the Lagrangian (\ref{LNCparticle})  has the form
\begin{equation}
\mL=\frac{1}{4\Gamma}\dot{X}^\mu \he_\mu^{ \ a}\delta_{ab}
\he_\nu^{ \ b}\dot{X}^\nu-2(m-q\varphi)^2\Phi\Gamma+q A_\mu\dot{X}^\mu \ .
\end{equation}
Finally we solve the equations of motion for $\Gamma$ and we get
\begin{equation}
\tau_\mu \dot{X}^\mu=2\Gamma(m-q\varphi)
\end{equation}
that implies $\Gamma=\frac{1}{2(K-q\varphi)}\dot{X}^\mu \tau_\mu
$ and hence the Lagrangian density has the form
\begin{eqnarray}\label{mLfinalNC}
\mL&=&\frac{(m-q\varphi)}{2\tau_\mu\dot{X}^\mu}\dot{X}^\mu
\he_\mu^{ \ a}\delta_{ab}\he_\nu^{ \ b}\dot{X}^\nu
-(m-q\varphi)\Phi\tau^\mu\dot{X}^\mu+aA_\mu\dot{X}^\mu=\nonumber \\
&=&\frac{(m-q\varphi)}{2\tau_\mu \dot{X}^\mu}\dot{X}^\mu
\bh_{\mu\nu}\dot{X}^\nu+q A_\mu\dot{X}^\mu \ ,
\nonumber \\
\end{eqnarray}
where in the last step we used the fact that
\begin{equation}
\he_\mu^{ \ a}\delta_{ab}\he_\nu^{ \ b}=\bh_{\mu\nu}+2\Phi\tau_\mu
\tau_\nu \ .
\end{equation}
This is an action for the massive particle in Newton-Cartan background and in
the background gauge field that coincides with the action found in
\cite{Festuccia:2016caf}.

Now we derive HJ equation for particle in NC background. We start with the
 conjugate momenta that follow from (\ref{mLfinalNC})
\begin{equation}\label{pmupartialS}
p_\mu=-\frac{(m-q\varphi)}{2(\tau_\mu\dot{X}^\mu)^2}\tau_\mu\dot{X}^\rho\bh_{\rho\sigma}
\dot{X}^\sigma+qA_\mu+\frac{(m-q\varphi)\bh_{\mu\nu}\dot{X}^\nu}{\tau_\rho
	\dot{X}^\rho}=\partial_\mu S \ .
\end{equation}
As the next step we
multiply this expression with $v^\mu$ and  we obtain
\begin{equation}
v^\mu(\partial_\mu S-q A_\mu+(m-q\varphi)M_\mu)=\frac{(m-q\varphi)}{2(\tau_\mu\dot{X}^\mu)^2}
\dot{X}^\rho h_{\rho\sigma}\dot{X}^\sigma
\end{equation}
On the other hand when we multiply (\ref{pmupartialS})  with $h^{\nu\mu}$ we get
\begin{equation}
h^{\nu\mu}(\partial_\mu S-qA_\mu)=
\frac{(m-q\varphi)}{\tau_\rho\dot{X}^\rho}h^{\nu\mu}h_{\mu\rho}\dot{X}^\rho
-(m-q\varphi)h^{\nu\mu}M_\mu \ .
\end{equation}
Collecting there results together we obtain following classical HJ equation for massive and charged particle in Newton-Cartan background
\begin{eqnarray}\label{HJNCparticle}
& &(\partial_\mu S-qA_\mu+(m-q\varphi)M_\mu)h^{\mu\nu}
(\partial_\nu S-qA_\nu+(m-q\varphi)M_\nu)-\nonumber \\
& & -2(m-q\varphi)v^\mu
(\partial_\mu S-qA_\mu+(m-q\varphi)M_\mu)=0 \ .
\nonumber \\
\end{eqnarray}
  For further purposes we rewrite this equation into the form
\begin{eqnarray}
& &2(m-q\varphi)\hv^\mu (\partial_\mu S-qA_\mu)-(\partial_\mu S-qA_\mu)h^{\mu\nu}(\partial_\nu S-qA_\nu)-2(m-q\varphi)^2\Phi=0
\ .
\nonumber \\
\end{eqnarray}
In order to find its quantum version we start with the action for complex scalar
field in NC background. We again derive this action using   null dimensional reduction.
More explicitly, let us consider  an action  for complex scalar field in
$D+1$ dimensions
\begin{equation}
I=\int d^{D}x\sqrt{-\gamma}
\left(-\gamma^{AB}D_A \Psi D_B\Psi^*\right) \ ,
\end{equation}
where $D_A=\partial_A \Psi-iq A_A\Psi$.
Now  we presume that the
background metric is given in
(\ref{gammametric}) and (\ref{gammametricinv}). We further presume
that all fields do not depend on $u$ and impose following ansatz for the  field $\Psi$
\begin{equation}
\Psi=e^{imu}\psi \ , \quad  \partial_u \Psi=
im \Psi \ .
\end{equation}
Then  we obtain an action for Schr\r{o}dinger
field in NC background
\begin{eqnarray}\label{SchNCb}
& &I=\int d^{D}xe( i(m-q\varphi)\hv^\mu(D_\mu\psi)^*\psi-i(m-q\varphi)\hv^\mu D_\mu\psi
\psi^*-\nonumber \\
& & -2\Phi (m-q\varphi)^2\psi \psi^*-h^{\mu\nu}D_\mu\psi (D_\nu\psi)^*) \ ,
\nonumber \\
\end{eqnarray}
where
\begin{equation}
D_\mu\psi=\partial_\mu\psi-iq A_\mu \psi \ .
\end{equation}
From (\ref{SchNCb})  we could find equations of motion for $\psi$ which are clearly linear.

As the next step we use polar parametrization of $\psi$ defined as
\begin{equation}
\psi=\sqrt{\rho}e^{iS} \ ,
\end{equation}
where $\rho$ and $S$ are real. Inserting this form of $\psi$ into
(\ref{SchNCb}) we obtain
\begin{eqnarray}\label{SchactSrho}
& &I^{sch}
=\int d^{D}x e\left(2(m-q\varphi)\rho \hv^\mu (\partial_\mu S-qA_\mu)
-2\Phi(m-q\varphi)^2\rho-\right.\nonumber \\
& &\left.-\frac{1}{4\rho}h^{\mu\nu}\partial_\mu\rho \partial_\nu\rho-
\rho h^{\mu\nu}(\partial_\mu S-qA_\mu)(\partial_\nu S-qA_\nu)\right) \ .
\nonumber \\
\end{eqnarray}
From (\ref{SchactSrho}) we obtain  following equations of motion
\begin{eqnarray}
& &\partial_\mu[e(m-q\varphi)\rho \hv^\mu]-\partial_\mu[e\rho h^{\mu\nu}
(\partial_\nu S-qA_\nu)]=0 \ , \nonumber \\
& &2e(m-q\varphi)\hv^\mu(\partial_\mu S-qA_\mu)-2e\Phi(m-q\varphi)^2+
\nonumber \\
& &+\frac{1}{4\rho^2}eh^{\mu\nu}\partial_\mu\rho\partial_\nu\rho+
\partial_\mu[\frac{1}{2\rho}eh^{\mu\nu}\partial_\nu\rho]
-eh^{\mu\nu}(\partial_\mu S-qA_\mu)(\partial_\nu S-qA_\nu)=0 \ .  \nonumber \\
\end{eqnarray}
The first equation can be interpreted as continuity equation  that expresses
conservation of the trajectories in the ensemble of  all possible trajectories of massive
particle in NC background while the second one can be written into the form
\begin{eqnarray}
& & 2(m-q\varphi)\hv^\mu(\partial_\mu S-qA_\mu)-h^{\mu\nu}(\partial_\mu S-qA_\mu)
(\partial_\nu S-qA_\nu)-\nonumber \\
& & -2\Phi (m-q\varphi)^2+Q=0 \ , \nonumber \\
\end{eqnarray}
where $Q$ is defined as
\begin{equation}
Q=\frac{1}{4\rho^2}h^{\mu\nu}\partial_\mu\rho \partial_\nu\rho+\frac{1}{2e}
\partial_\mu\left[\frac{1}{\rho}eh^{\mu\nu}\partial_\nu\rho \right] \ .
\end{equation}
It is clear that classical action corresponds to the situation when $Q=0$. Then the equations of motion for $S$ and $\rho$ can be derived from the action
\begin{equation}
I^{sch}_{clas}=\int d^{D}x
e(2(m-q\varphi)\rho \hv^\mu (\partial_\mu S-qA_\mu)-2\Phi (m-q\varphi)^2\rho
-\rho h^{\mu\nu}(\partial_\mu S-qA_\mu)(\partial_\nu S-qA_\nu)) \
\end{equation}
or equivalently using complex field $\psi_{clas}=\sqrt{\rho}e^{iS}$
\begin{eqnarray}
& &I^{sch}_{clas}=\int d^{D}x e\left(i(m-q\varphi)\hv^\mu (\psi_{clas} (D_\mu\psi_{clas})^*-\psi^*_{clas}
D_\mu\psi_{clas})-2\Phi (m-q\varphi)^2\psi_{clas} \psi^*_{clas} \right.\nonumber \\
& & \left.-\frac{1}{2}h^{\mu\nu}
D_\mu\psi_{clas} (D_\nu\psi_{clas})^*+
 +\frac{1}{4}h^{\mu\nu}\frac{D_\mu\psi_{clas} D_\nu\psi_{clas}}{\psi_{clas}\psi_{clas}}+
\frac{1}{4}h^{\mu\nu}\frac{(D_\mu\psi_{clas})^*(D_\nu\psi_{clas})^*}{\psi^*_{clas}
\psi^*_{clas}}\right) \ .  \nonumber \\
\end{eqnarray}
We again see that resulting equations of motions will be non-linear due to the
last two terms on the second line and hence superposition of two solutions of the equations of motion of classical particle in NC background is not solutions of these equations.

Finally we will try to find quantum version of Lagrangian for massive particle
in NC background starting with the quantum HJ equation. Replacing $\partial_\mu S$ with $p_\mu$ we find that it corresponds to the Hamiltonian constraint in the form
\begin{eqnarray}
& & \mH_\tau^{q}=
2(m-q\varphi)\hv^\mu(p_\mu-qA_\mu)-h^{\mu\nu}(p_\mu-qA_\mu)
(p_\nu-qA_\nu)-\nonumber \\
& &-2\Phi (m-q\varphi)^2+Q=0 \  \nonumber \\
\end{eqnarray}
so that the Hamiltonian is $H^q=\Gamma \mH_\tau^q$,  where $\Gamma$ is corresponding Lagrange multiplier. With the help of this Hamiltonian
we find
\begin{equation}
\dot{X}^\mu=\pb{X^\mu,H^q}=2\Gamma ((m-q\varphi)\hv^\mu-h^{\mu\nu}(p_\nu-qA_\nu))
\end{equation}
and hence
\begin{eqnarray}
L^q=p_\mu\dot{X}^\mu-H^q
=\frac{(m-q\varphi)}{2\dot{X}^\mu\tau_\mu}\dot{X}^\mu\bh_{\mu\nu}\dot{X}^\nu+qA_\mu+
\frac{1}{m-q\varphi}Q\tau_\mu\dot{X}^\mu \
\end{eqnarray}
using the fact that from the equation of motion for $X^\mu$ we obtain
\begin{equation}
\Gamma=-\frac{\dot{X}^\mu\tau_\mu}{2(m-q\varphi)} \ .
\end{equation}
We see fundamental difference between relativistic quantum particle and quantum particle in NC background which is in the presence of the quantum potential in both Lagrangians where in case of Newton-Cartan quantum particle the quantum potential is simply added to the Lagrangian of classical particle while in case of quantum relativistic particle it appears under square root together with $m^2$.
\section{Conclusion}\label{fifth}
In this short note we performed analysis of classical and quantum particles in relativistic and Newton-Cartan background in the framework of de Broglie-Bohm quantum mechanics. We determined classical and quantum version of HJ equation for relativistic particle and discuss properties of the wave function description of both theories. We performed the same analysis in case of particle in Newton-Cartan background where we determined classical and quantum HJ equation. We mean that this is really interesting result since we found covariant version of non-relativistic de-Broglie-Bohm quantum mechanics.

De Broglie-Bohm formulation of quantum mechanics is very fascinating and certainly deserves further study and analysis. In particular, it is very interesting that Lagrangian for quantum particle contains quantum potential that depends on derivatives of density of trajectories at space-time point. Further,
it can be shown that the Schrödinger equation
may be recast as a self-contained second-order Newtonian law for a
congruence of spacetime trajectories, see for example
\cite{Holland2}. It would be certainly very interesting to perform similar analysis in case of massive particle in Newton-Cartan background. We hope to return to this problem in near future.

 {\bf Acknowledgement:}

 This  work  was
supported by the Grant Agency of the Czech Republic under the grant
P201/12/G028.
%
%


\end{document}